\documentclass[sigplan,screen,natbib=false]{acmart}

\AtBeginDocument{%
  }

\setcopyright{acmcopyright}
\copyrightyear{2023}
\acmYear{2023}
\acmDOI{XXXXXXX.XXXXXXX}

\acmConference[IWACO '23]{of the 9th International Workshop on Aliasing, Confinement and Ownership}{Sun 22 - Fri 27 October, 2023}{Cascais, Portugal}
\RequirePackage[
  datamodel=acmdatamodel,
  style=acmauthoryear,
]{biblatex}

\RequirePackage{enumitem}
\RequirePackage{xspace}
\RequirePackage{ifthen}
\RequirePackage{xcolor}
\RequirePackage{caption}
\RequirePackage{float}
\RequirePackage{mathpartir}
\RequirePackage{graphicx}
\RequirePackage{namespc}
\RequirePackage[frozencache=true,cachedir=minted-cache]{minted}
\RequirePackage{csquotes}
\RequirePackage{tikz}
\usetikzlibrary{positioning,fit}

\newcommand{\boxcalc}{box calculus\xspace}
\newcommand{\cfsub}{System \texorpdfstring{$\textrm{CF}_\sub$\xspace}{CF<:}}
\newcommand{\ccsubbox}{System \texorpdfstring{$\textrm{CC}_{\sub\square}$\xspace}{CC<:□}}

\newcommand{\fsub}{System \texorpdfstring{$\textrm{F}_\sub$\xspace}{F<:}}
\newcommand{\poplmark}{\textsc{POPLMark}\xspace}
\newcommand{\autosubst}{\textsc{AutoSubst}\xspace}

\preto\tabular{\setcounter{magicrownumbers}{0}}
\newcounter{magicrownumbers}

\renewcommand{\texttilde}{\raisebox{0.5ex}{\texttildelow}}
\newcommand{\sep}{\;\mid\;}

\newcounter{synconindex}
\newenvironment{syncat}[1]{
    \setcounter{synconindex}{0}
    \begin{array}{lcllcll}
        #1
}{
    \end{array}
}

\newcommand{\syncon}[1]{
    &
    \ifthenelse{\equal{\thesynconindex}{0}}{\Coloneqq}{\sep}
    & #1
    \stepcounter{synconindex}
}

\newcounter{termindent}

\ExplSyntaxOn
\NewExpandableDocumentCommand{\repeatn}{O{}mm}
 {
  \int_compare:nT { #2 > 0 }
   {
    #3 \prg_replicate:nn { #2 - 1 } { #1#3 }
   }
 }
\ExplSyntaxOff

\newcommand{\rulename}[1]{\textsc{\small (#1)}}
\newcommand{\ir}[3]{\inferrule{#2}{#3} \; \rulename{#1}}

\newcommand{\keyword}[1]{\ensuremath{\mathbf{#1}}}

\newcommand{\sym}[1]{\color[HTML]{000000}#1\xspace}
\newcommand{\var}[1]{\ensuremath{{\color[HTML]{000000}#1}}}
\newcommand{\bvar}[1]{\ensuremath{{\color[HTML]{ff0000}#1}}}
\newcommand{\fvar}[1]{\ensuremath{{\color[HTML]{008800}#1}}}
\newcommand{\tvar}[1]{\ty{#1}}
\newcommand{\ty}[1]{\ensuremath{{\color[HTML]{1111ff}#1}}}
\newcommand{\prety}[1]{\ensuremath{{\color[HTML]{111188}#1}}}
\newcommand{\ctx}[1]{\ty{#1}}
\renewcommand{\cap}[1]{\ensuremath{{\color[HTML]{008800}#1}}}
\newcommand{\tm}[1]{\ensuremath{{\color[HTML]{000000}#1}}}
\newcommand{\storectx}[1]{\ensuremath{{\color[HTML]{000000}#1}}}
\newcommand{\evalctx}[1]{\ensuremath{{\color[HTML]{000000}#1}}}

\newcommand{\sub}{{\ensuremath \sym{<:}}}

\newcommand{\dom}[1]{\ensuremath{\color[HTML]{000000}\mathbf{dom}(\ctx{#1})}}

\newcommand{\subtyping}[3]{\ensuremath{\ctx{#1} \sym{\vdash} \ty{#2} \; \sub \; \ty{#3}}}
\newcommand{\subcapt}[3]{\ensuremath{\ctx{#1} \sym{\vdash} \cap{#2} \; \sub_C \; \cap{#3}}}

\newcommand{\type}[1]{\ensuremath{\ty{#1} \; \textbf{type}}}
\newcommand{\pure}[1]{\ensuremath{\prety{#1} \; \textbf{pure}}}
\newcommand{\capt}[1]{\ensuremath{\cap{#1} \; \textbf{capt}}}
\newcommand{\wfenv}[1]{\ensuremath{\ctx{#1} \; \textbf{wf}}}

\namespace{standard-fsub}{
    \providecommand{\tytop}{\ensuremath{\ty{\top}}}
    \providecommand{\tyvar}[1]{\ensuremath{\ty{#1}}}
    \providecommand{\tyarr}[2]{\ensuremath{\ty{#1 \to #2}}}
    \providecommand{\tyall}[2]{\ensuremath{\sym{\forall} \ty{#1} \to \ty{#2}}}
    
    \providecommand{\tmabs}[3]{\ensuremath{\tm{\sym{\lambda} \sym{(}\btyp{#1}{#2}\sym{)} \sym{.} \; #3}}}
    \providecommand{\tmtabs}[3]{\ensuremath{\tm{\sym{\Lambda} \sym{[}\bsub{#1}{#2}\sym{]} \sym{.} \; #3}}}
    \providecommand{\tmapp}[2]{\ensuremath{\tm{#1} \tm{#2}}}
    \providecommand{\tmtapp}[2]{\ensuremath{\tm{#1} \sym{[}\ty{#2}\sym{]}}}
    \providecommand{\tmlet}[3]{\ensuremath{\keyword{let} \; \var{#1} = \tm{#2} \; \keyword{in} \; \tm{#3}}}

    \providecommand{\btyp}[2]{\ensuremath{\var{#1} \sym{:} \ty{#2}}}
    \providecommand{\bsub}[2]{\ensuremath{\var{#1} \; \sym{\sub} \; \ty{#2}}}
    \providecommand{\ctxtyp}[3]{\ensuremath{\ctx{#1}\sym{,} \btyp{#2}{#3}}}
    \providecommand{\ctxsub}[3]{\ensuremath{\ctx{#1}\sym{,} \bsub{#2}{#3}}}
    
    \providecommand{\lnopen}[3]{\ensuremath{\sym{[}\bvar{#1} \sym{\to} \tm{#2}\sym{]} \tm{#3}}}
    \providecommand{\lnsubst}[3]{\ensuremath{\sym{[}\fvar{#1} \sym{\to} \tm{#2}\sym{]} \tm{#3}}}
    \providecommand{\lnsubstt}[3]{\ensuremath{\sym{\{}\tvar{#1} \sym{\to} \ty{#2}\sym{\}} \tm{#3}}}
    \providecommand{\storecons}[3]{\ensuremath{\storectx{#1} \sym{,} \var{#2} \sym{=} \tm{#3}}}

    \providecommand{\evalcons}[2]{\ensuremath{\var{#1} \sym{::} \evalctx{#2}}}

    \providecommand{\mkstate}[3]{\ensuremath{\langle #1 \mid #2 \mid \tm{#3} \rangle}}

    \providecommand{\red}[3]{\ensuremath{#2 \xrightarrow{\fvar{#1}} #3}}

    \newcommand{\wftyp}[2]{\ensuremath{\ctx{#1} \sym{\vdash} \ty{#2} \; \textbf{wf}}}
}{}

\namespace{locallynameless}{
    \providecommand{\tytop}{\ensuremath{\ty{\top}}}
    \providecommand{\tyvar}[1]{\ensuremath{\ty{#1}}}
    \providecommand{\tyarr}[2]{\ensuremath{\ty{#1 \to #2}}}
    \providecommand{\tmabs}[2]{\ensuremath{\tm{\sym{\lambda} \sym{(}\ty{#1}\sym{)} #2}}}
    \providecommand{\tmtabs}[2]{\ensuremath{\tm{\sym{\Lambda} \sym{[}\ty{#1}\sym{]} #2}}}
    \providecommand{\tmapp}[2]{\ensuremath{\tm{#1} \tm{#2}}}
    \providecommand{\btyp}[2]{\ensuremath{\var{#1} \sym{:} \ty{#2}}}
    \providecommand{\bsub}[2]{\ensuremath{\var{#1} \; \sym{\sub} \; \ty{#2}}}
    \providecommand{\ctxtyp}[3]{\ensuremath{\ctx{#1}\sym{,} \btyp{#2}{#3}}}
    \providecommand{\ctxsub}[3]{\ensuremath{\ctx{#1}\sym{,} \bsub{#2}{#3}}}
    
    \providecommand{\lnopen}[3]{\ensuremath{\sym{[}\bvar{#1} \sym{\to} \tm{#2}\sym{]} \tm{#3}}}

    \newcommand{\wftyp}[2]{\ensuremath{\ctx{#1} \sym{\vdash} \ty{#2} \; \textbf{wf}}}
}{}

\namespace{anf-fsub}{
    \providecommand{\tytop}{\ensuremath{\ty{\top}}}
    \providecommand{\tyvar}[1]{\ensuremath{\ty{#1}}}
    \providecommand{\tyarr}[2]{\ensuremath{\ty{#1} \to \ty{#2}}}
    \providecommand{\tyall}[2]{\ensuremath{\sym{\forall} \ty{#1} \to \ty{#2}}}
    
    \providecommand{\tmabs}[2]{\ensuremath{\tm{\sym{\lambda} \sym{(}\ty{#1}\sym{)} #2}}}
    \providecommand{\tmtabs}[2]{\ensuremath{\tm{\sym{\Lambda} \sym{[}\ty{#1}\sym{]} #2}}}
    \providecommand{\tmapp}[2]{\ensuremath{\var{#1} \var{#2}}}
    \providecommand{\tmtapp}[2]{\ensuremath{\tm{#1} \sym{[}\ty{#2}\sym{]}}}
    \providecommand{\tmlet}[2]{\ensuremath{\keyword{let} \; \var{#1} \; \keyword{in} \; \tm{#2}}}

    \providecommand{\btyp}[2]{\ensuremath{\var{#1} \sym{:} \ty{#2}}}
    \providecommand{\bsub}[2]{\ensuremath{\var{#1} \; \sym{\sub} \; \ty{#2}}}
    \providecommand{\ctxtyp}[3]{\ensuremath{\ctx{#1}\sym{,} \btyp{#2}{#3}}}
    \providecommand{\ctxsub}[3]{\ensuremath{\ctx{#1}\sym{,} \bsub{#2}{#3}}}
    
    \providecommand{\lnopen}[3]{\ensuremath{\sym{[}\bvar{#1} \sym{\to} \tm{#2}\sym{]} \tm{#3}}}
    \providecommand{\lnopent}[3]{\ensuremath{\sym{\{}\bvar{#1} \sym{\to} \ty{#2}\sym{\}} \tm{#3}}}
    \providecommand{\lnsubst}[3]{\ensuremath{\sym{[}\fvar{#1} \sym{\to} \tm{#2}\sym{]} \tm{#3}}}
    \providecommand{\lnsubstt}[3]{\ensuremath{\sym{\{}\tvar{#1} \sym{\to} \ty{#2}\sym{\}} \tm{#3}}}
    \providecommand{\storecons}[3]{\ensuremath{\storectx{#1} \sym{,} \var{#2} \sym{=} \tm{#3}}}

    \providecommand{\evalcons}[2]{\ensuremath{\var{#1} \sym{::} \evalctx{#2}}}

    \providecommand{\mkstate}[3]{\ensuremath{\langle #1 \mid #2 \mid \tm{#3} \rangle}}

    \providecommand{\red}[3]{\ensuremath{#2 \xrightarrow{\fvar{#1}} #3}}

    \newcommand{\wftyp}[2]{\ensuremath{\ctx{#1} \sym{\vdash} \ty{#2} \; \textbf{wf}}}
}{}

\namespace{anf-cfsub}{
    \providecommand{\tyvar}[1]{\ensuremath{\ty{#1}}}
    \providecommand{\tytop}{\ensuremath{\prety{\top}}}
    \providecommand{\tyarr}[2]{\ensuremath{\prety{\sym{\forall} \sym{(} \ty{#1} \sym{)} \sym{\to} \ty{#2}}}}
    \providecommand{\tyall}[2]{\ensuremath{\prety{\sym{\forall} \sym{[} \ty{#1} \sym{]} \sym{\to} \ty{#2}}}}
    \providecommand{\tycapt}[2]{\ensuremath{\cap{#1} \prety{#2}}}
    \providecommand{\tycaptc}[2]{\ensuremath{\tycapt{\sym{\{} \var{#1} \sym{\}}}{#2}}}
    \providecommand{\tmabs}[2]{\ensuremath{\tm{\sym{\lambda} \sym{(}\ty{#1}\sym{)} #2}}}
    \providecommand{\tmtabs}[2]{\ensuremath{\tm{\sym{\Lambda} \sym{[}\ty{#1}\sym{]} #2}}}
    \providecommand{\tmapp}[2]{\ensuremath{\fvar{#1} \fvar{#2}}}
    \providecommand{\tmtapp}[2]{\ensuremath{\tm{#1} \sym{[}\ty{#2}\sym{]}}}
    \providecommand{\tmlet}[2]{\ensuremath{\keyword{let} \; \var{#1} \; \keyword{in} \; \tm{#2}}}

    \providecommand{\btyp}[2]{\ensuremath{\var{#1} \sym{:} \ty{#2}}}
    \providecommand{\bsub}[2]{\ensuremath{\var{#1} \; \sym{\sub} \; \ty{#2}}}
    \providecommand{\ctxtyp}[3]{\ensuremath{\ctx{#1}\sym{,} \btyp{#2}{#3}}}
    \providecommand{\ctxsub}[3]{\ensuremath{\ctx{#1}\sym{,} \bsub{#2}{#3}}}
    
    \providecommand{\lnopen}[3]{\ensuremath{\sym{[}\bvar{#1} \sym{\to} \tm{#2} \sym{]} \tm{#3}}}
    \providecommand{\lnopent}[3]{\ensuremath{\sym{\{}\bvar{#1} \sym{\to} \ty{#2} \sym{\}} \tm{#3}}}
    \providecommand{\lnopenct}[3]{\ensuremath{\sym{\{}\bvar{#1} \sym{\to} #2 \sym{\}} \tm{#3}}}
    \providecommand{\lnsubst}[3]{\ensuremath{\sym{[}\fvar{#1} \sym{\to} \tm{#2}\sym{]} \tm{#3}}}
    \providecommand{\lnsubstt}[3]{\ensuremath{\sym{\{}\tvar{#1} \sym{\to} \ty{#2}\sym{\}} \tm{#3}}}
    \providecommand{\storecons}[3]{\ensuremath{\storectx{#1} \sym{,} \var{#2} \sym{=} \tm{#3}}}

    \providecommand{\evalcons}[2]{\ensuremath{\var{#1} \sym{::} \evalctx{#2}}}

    \providecommand{\mkstate}[3]{\ensuremath{\langle #1 \mid #2 \mid \tm{#3} \rangle}}

    \providecommand{\red}[3]{\ensuremath{#2 \xrightarrow{\fvar{#1}} #3}}

    \newcommand{\wftypin}[4]{\ensuremath{\ctx{#1} ; \fvar{#2} ; \fvar{#3} \sym{\vdash_T} \ty{#4} \; \textbf{wf}}}
    \newcommand{\wfpretypin}[4]{\ensuremath{\ctx{#1} ; \fvar{#2} ; \fvar{#3} \sym{\vdash_R} \prety{#4} \; \textbf{wf}}}

    \newcommand{\wftyp}[2]{\ensuremath{\ctx{#1} \sym{\vdash}_T \ty{#2} \; \textbf{wf}}}
    \newcommand{\wfpretyp}[2]{\ensuremath{\ctx{#1} \sym{\vdash}_R \prety{#2} \; \textbf{wf}}}

    \newcommand{\subtyp}[3]{\ensuremath{\ctx{#1} \sym{\vdash}_T \ty{#2} \; \sub \; \ty{#3}}}
    \newcommand{\subpretyp}[3]{\ensuremath{\ctx{#1} \sym{\vdash}_R \prety{#2} \; \sub \; \prety{#3}}}
}{}

\namespace{anf-ccsubbox}{
    \providecommand{\tyvar}[1]{\ensuremath{\ty{#1}}}
    \providecommand{\tytop}{\ensuremath{\prety{\top}}}
    \providecommand{\tyarr}[2]{\ensuremath{\prety{\sym{\forall} \sym{(} \ty{#1} \sym{)} \sym{\to} \ty{#2}}}}
    \providecommand{\tyall}[2]{\ensuremath{\prety{\sym{\forall} \sym{[} \prety{#1} \sym{]} \sym{\to} \ty{#2}}}}
    \providecommand{\tycapt}[2]{\ensuremath{\cap{#1} \prety{#2}}}
    \providecommand{\tybox}[1]{\ensuremath{\sym{\square} \ty{#1}}}
    \providecommand{\tycaptc}[2]{\ensuremath{\tycapt{\sym{\{} \var{#1} \sym{\}}}{#2}}}
    \providecommand{\tmabs}[2]{\ensuremath{\tm{\sym{\lambda} \sym{(}\ty{#1}\sym{)} #2}}}
    \providecommand{\tmtabs}[2]{\ensuremath{\tm{\sym{\Lambda} \sym{[}\ty{#1}\sym{]} #2}}}
    \providecommand{\tmapp}[2]{\ensuremath{\fvar{#1} \fvar{#2}}}
    \providecommand{\tmtapp}[2]{\ensuremath{\tm{#1} \sym{[}\ty{#2}\sym{]}}}
    \providecommand{\tmlet}[2]{\ensuremath{\keyword{let} \; \var{#1} \; \keyword{in} \; \tm{#2}}}
    \providecommand{\tmunbox}[2]{\ensuremath{\cap{#1} \sym{\multimapinv}} \fvar{x}}
    \providecommand{\tmbox}[1]{\ensuremath{\sym{\square} \fvar{#1}}}

    \providecommand{\btyp}[2]{\ensuremath{\var{#1} \sym{:} \ty{#2}}}
    \providecommand{\bsub}[2]{\ensuremath{\var{#1} \; \sym{\sub} \; \ty{#2}}}
    \providecommand{\ctxtyp}[3]{\ensuremath{\ctx{#1}\sym{,} \btyp{#2}{#3}}}
    \providecommand{\ctxsub}[3]{\ensuremath{\ctx{#1}\sym{,} \bsub{#2}{#3}}}
    
    \providecommand{\lnopen}[3]{\ensuremath{\sym{[}\bvar{#1} \sym{\to} \tm{#2} \sym{]} \tm{#3}}}
    \providecommand{\lnopent}[3]{\ensuremath{\sym{\{}\bvar{#1} \sym{\to} \ty{#2} \sym{\}} \tm{#3}}}
    \providecommand{\lnopenct}[3]{\ensuremath{\sym{\{}\bvar{#1} \sym{\to} #2 \sym{\}} \tm{#3}}}
    \providecommand{\lnsubst}[3]{\ensuremath{\sym{[}\fvar{#1} \sym{\to} \tm{#2}\sym{]} \tm{#3}}}
    \providecommand{\lnsubstt}[3]{\ensuremath{\sym{\{}\tvar{#1} \sym{\to} \ty{#2}\sym{\}} \tm{#3}}}
    \providecommand{\storecons}[3]{\ensuremath{\storectx{#1} \sym{,} \var{#2} \sym{=} \tm{#3}}}

    \providecommand{\evalcons}[2]{\ensuremath{\var{#1} \sym{::} \evalctx{#2}}}

    \providecommand{\mkstate}[3]{\ensuremath{\langle #1 \mid #2 \mid \tm{#3} \rangle}}

    \providecommand{\red}[3]{\ensuremath{#2 \xrightarrow{\fvar{#1}} #3}}

    \newcommand{\wftyp}[2]{\ensuremath{\ctx{#1} \sym{\vdash} \ty{#2} \; \textbf{wf}}}

}{}

\begin{document}

\title{A Mechanized Theory of the Box Calculus}

\author{Joseph Fourment}
\email{joseph.fourment@epfl.ch}
\orcid{0009-0009-5982-3670}
\affiliation{%
  \institution{EPFL}
  \streetaddress{P.O. Box 1212}
  \city{Lausanne}
  \country{Switzerland}
  \postcode{1024}
}

\author{Yichen Xu}
\email{yichen.xu@epfl.ch}
\orcid{0000-0003-2089-6767}
\affiliation{%
  \institution{EPFL}
  \streetaddress{P.O. Box 1212}
  \city{Lausanne}
  \country{Switzerland}
  \postcode{1024}
}

\renewcommand{\shortauthors}{Fourment and Xu}

\begin{abstract}
    The capture calculus is an extension of \fsub that tracks free variables of terms in their type, allowing one to represent capabilities while limiting their scope.
    While previous calculi had mechanized soundness proofs --- notably \cfsub --- the latest version, namely the box calculus (\ccsubbox), only had a paper proof.
    We present here our work on mechanizing the theory of the box calculus in Coq, and the challenges encountered along the way.
    While doing so, we motivate the current design of capture calculus, in particular the concept of \textit{boxes}, from both user and metatheoretical standpoints.
    Our mechanization is complete and \href{https://github.com/felko/ccsubbox}{available on GitHub}.
\end{abstract}



%%  \includegraphics[width=\textwidth]{sampleteaser}
%%  \caption{Seattle Mariners at Spring Training, 2010.}
%%  \Description{Enjoying the baseball game from the third-base
%%  seats. Ichiro Suzuki preparing to bat.}

\received{12 July 2023}
%\received[revised]{12 July 2023}
%\received[accepted]{12 July 2023}


\maketitle

    \section{Introduction}

Capture checking is an experimental Scala feature that aims to provide a basis
for new type system abilities, such as checked exceptions --- particularly in presence of higher-order functions --- \cite{odersky2021safer}, algebraic effects \cite{plotkin2009handlers}, and safe memory management through regions \cite{tofte1997region}.
It does so by exposing type level information about free variables in terms.

To investigate the metatheory of capture checking, two calculi have been introduced, namely \cfsub \cite{boruch2021tracking} and later the \boxcalc (\ccsubbox) \cite{odersky2022scoped}.
While \cfsub is fully mechanized in Coq, the soundness proofs of other variants of
the calculus are not yet mechanized.
In particular, \ccsubbox differs from the initial \cfsub in that it uses monadic normal form (MNF) syntax \cite{hatcliff1994generic}.
Switching to MNF has several prospects regarding the formalization of a larger fragment of Scala, notably path-dependent types \cite{rapoport2019path}.

Another difference is that \ccsubbox restricts type abstractions and type applications to \textit{pure} (uncaptured) types.
To recover the unrestricted type abstraction, one uses \textit{boxing} which consists in hiding the capturing type behind a new modality denoted $\square$.
The seemingly simple restriction on type abstractions drastically simplifies the metatheory both in prosaic and mechanized proof as we will show.
Our work is composed of multiple steps illustrated in Figure~\ref{fig:roadmap}.
This paper mostly focuses on the mechanization of \ccsubbox, the mechanizations of \href{https://github.com/felko/anf-fsub}{MNF-\fsub} and \href{https://github.com/felko/ccsubbox/tree/mnf-cfsub}{MNF-\cfsub} are also available on GitHub.

\begin{figure}[h]
    \resizebox{\columnwidth}{!}{%
        \begin{tikzpicture}[text centered]
    	\node (fsub)      at (0, 1.5) {\begin{tabular}{c} \fsub \\ \footnotesize \cite{aydemir2008engineering} \end{tabular}};
            \node (cfsub)     at (3, 1.5) {\begin{tabular}{c} \cfsub \\ \footnotesize (Edward Lee) \end{tabular}};
            \node (mnf-fsub)  at (0, 0) {MNF-\fsub};
            \node (mnf-cfsub) at (3, 0) {MNF-\cfsub};
            \node (ccsubbox)  at (6, 0) {\ccsubbox};
    
            \draw[->] (fsub) -- (cfsub);
            \draw[->] (fsub) -- (mnf-fsub);
            \draw[->] (cfsub) -- (mnf-cfsub);
            \draw[->] (mnf-fsub) -- (mnf-cfsub);
            \draw[->] (mnf-cfsub) -- (ccsubbox);
    
            \node (box) [draw=red,minimum height=3em,rounded corners,fit=(mnf-fsub) (mnf-cfsub) (ccsubbox)] {};
            \node [above left=.5em of box.north east,red] {this work};
        \end{tikzpicture}%
    }%
    \caption{Roadmap for the mechanization of \ccsubbox}%
    \label{fig:roadmap}%
\end{figure}%

    \section{Motivation}

Here we explain the rationale behind the capture calculus and what problem it is trying to solve, illustrating its purpose with concrete use cases expressed in Scala.
We take inspiration in the examples detailed in \cite{odersky2022scoped}.

        \subsection{Capabilities \& Monadic Reflection}

Monadic effects are currently the standard tool to deal with effects in functional programming.
However, monads notoriously don't compose in general, we need extra machinery to compose them when it is legal to do so, such as \textit{monad transformers}.
A more recent approach is the theory of algebraic effects with handlers \cite{plotkin2009handlers}, which restricts the expressible effects to a subset of so-called algebraic effects, that compose out of the box.
This approach has been implemented as libraries, e.g. in Haskell \cite{kiselyov2015freer}, using an integrated ambient monad parametrized with the kind of effects that a computation is able to perform.

One drawback with any kind of monadic effect system is that effectful computations are expressed in continuation-passing style.
While syntactic sugar such as Haskell's \mintinline{haskell}|do|-notation or Scala's \mintinline{scala}|for-yield| syntax can make such code look like direct style, arbitrary control flow structures such as \mintinline{scala}|while| loops are not permitted.
Moreover, monadic code often introduces some overhead due to the allocation of closures to represent the continuations, and due to the additional calls, both to the binding operation (\mintinline{haskell}|>>=| or \mintinline{scala}|flatMap|) and the continuations.

Ideally, we would like to write effectful functions in direct-style.
To this end, algebraic effects have also been implemented in standalone compilers, with their semantics based on delimited continuations.
However, another approach is to use \textit{monadic reflection} \cite{filinski1994representing, filinsky1999layered}, which allows us to use monads in direct-style.
Monadic reflection can be implemented on top of Scala \cite{brachthauser2021representing} provided an implementation of delimited continuations.
In particular, delimited continuations can be implemented on top of coroutines, of which \href{https://openjdk.org/projects/loom/}{Project Loom}, a fork of OpenJDK that aims to implement lightweight threading, has built-in support.
Using monadic reflection, effectful computations can be encoded in Scala using \textit{context functions}, taking the context argument as the capability.
A capability does not have any existence at runtime, it is merely a permission to perform certain functions.
For example, a function \mintinline{scala}|f| that can fail with an error of type \mintinline{scala}|E| and can return a \mintinline{scala}|T| will be typed \mintinline{scala}|f: CanThrow[E] ?=> T| and can be declared as \\ \mintinline{scala}|def f(using err: CanThrow[E]): T|.
Context functions will resolve the capability with the \mintinline{scala}|given| mechanism.

        \subsection{Example}

\textit{Error handling.} The \mintinline{scala}|try| block can be used to generate a \mintinline{scala}|CanThrow| capability.

\begin{minted}[fontsize=\small]{scala}
def sqrt(x: Double)
        (using err: CanThrow[NegSqrt]): Double =
  if x < 0 then err.throw(NegSqrt(x)) else x ^ 0.5

def sqrtThunk(x: Double): () => Option[Double] =
  try () => Some(sqrt(x))
  catch case exc: NegSqrt => (() => None)
\end{minted}

At the beginning of the \mintinline{scala}|try| block, the compiler will insert the value \mintinline{scala}|val err: CanThrow[NegSqrt] = ???|.
Note that its definition is unimportant, as capabilities are erased at runtime.

However, there is a pitfall in this approach.
Consider now the value \mintinline{scala}|val f = sqrtThunk(-1)|. The function \mintinline{scala}|f| has type \mintinline[breaklines]{scala}|() => Option[Double]| but \mintinline{scala}|f()| will throw an uncaught error, despite the surrounding \mintinline{scala}|try| block, because the handler is no longer in scope when the thunk is called.

\textit{Resource management.} Consider a function \mintinline{scala}|withFile| defined below:

\begin{minted}[fontsize=\small]{scala}
def withFile[T](p: String, f: OutputStream => T): T =
    val file = new FileOutputStream(p)
    val result = f(file)
    file.close()
    result
\end{minted}

The function \mintinline{scala}|withFile| allows one to temporarily open a file, write in it, and then finally close the file.
This is analogous with the \mintinline{scala}|try|-with-resources idiom.
However, consider a function that creates a logger, returning a closure that contains the temporary stream:

\begin{minted}[fontsize=\small]{scala}
def makeLogger(p: String): String => () =
    withFile(p, stream => str => stream.write(str))
\end{minted}

Calling the resulting closure of \mintinline{scala}|makeLogger| is ill-behaved, because it is executed after the stream is closed.

        \subsection{The Capture Calculus}

\begin{anf-cfsub}

The purpose of the capture calculus is to prevent cases such as \mintinline{scala}|makeLogger| by statically preventing scoped capabilities to leak in closures.
It does so by enabling the types to represent which free variables are captured in terms.
As the capture calculus is meant to be integrated in Scala, a functional language with subtyping, it is based on \fsub.
The main addition of the capture calculus compared to \fsub is the notion of a \textit{capture set}.
A capture set is a finite set of variables attached to a type, such a type is called a \textit{capturing type}, and is denoted $\tycapt{C}{R}$ where $\cap{C}$ is the capture set and $\prety{R}$ is the underlying type which must be uncaptured (we say that $\prety{R}$ is a \textit{pure} type).
For example, the type $\tycapt{\{a, b, c\}}{\prety{\textbf{Int}}}$ is the type of terms of pure type $\prety{\textbf{Int}}$ that can mention $a$, $b$ and $c$ as free variables and no more --- but potentially less due to subtyping, or more precisely \textit{subcapturing}.

The subcapturing relation defines a partial order on capture set.
The relation is analogous to the powerset poset ordered by inclusion, with the important difference that variables present in a capture set can be expanded into their own capture set with a combination of the \rulename{sc-set} and \rulename{sc-var} rules written below:

\begin{mathpar}
    \ir{sc-set}{
        \subcapt{\Gamma}{\{x_1\}}{C} \and
        ... \and
        \subcapt{\Gamma}{\{x_n\}}{C}
    }{
        \subcapt{\Gamma}{\{x_1, ..., x_n\}}{C}
    } \\
    \ir{sc-var}{
        \btyp{x}{\tycapt{C_1}{R}} \sym{\in} \ctx{\Gamma} \and
        \subcapt{\Gamma}{C_1}{C_2}
    }{
        \subcapt{\Gamma}{\{x\}}{C_2}
    }
\end{mathpar}
\end{anf-cfsub}

\begin{anf-ccsubbox}
Note that, as variables can occur in types, the capture calculus is dependent, in a weak sense.
More precisely, we want to allow the return type of a function to reference the name of its parameter, so instead of having the usual function type as $\ty{T} \to \ty{U}$, function types are denoted $\tyarr{\var{x} \sym{:} \ty{T}}{U}$, where $\var{x}$ can be referenced in a capture set of $\ty{U}$.
%Formally, the well-formedness rule for function types is:
%\begin{mathpar}
%    \ir{fun-wf}{
%        \wftyp{\Gamma}{T} \and
%        \wftyp{\ctxtyp{\Gamma}{x}{T}}{U}
%    }{
%        \wftyp{\Gamma}{\tyarr{\var{x} \sym{:} \ty{T}}{U}}
%    }
%\end{mathpar}
\end{anf-ccsubbox}

Going back to our \mintinline{scala}|sqrt| example, we can no longer type \mintinline{scala}|sqrtThunk|.
Informally, this is because the resulting closure now has type \\ \mintinline{scala}|{err} (() => Option[Double])| which cannot be returned as it captures a local capability.
Note that the closure does not need to capture \mintinline{scala}|x| because of the \rulename{sc-var} rule of the subcapturing judgement, as the capture set of \mintinline{scala}|x| itself is empty.

\begin{minted}[fontsize=\small,escapeinside=||,mathescape=true]{scala}
def sqrtThunk(x: Double) =
  try
    val err: CanThrow[NegSqrt] = ???
    () => Some(sqrt(x)(using err))
    // has type {err} (() => Option[Double])
  catch case exc: NegSqrt => () => None
\end{minted}

        \subsection{Effect Polymorphism}

A typical challenge of effect systems is the notion of \textit{effect polymorphism}.
In presence of higher-order functions, we often want to allow the parameter function to be effectful.
Most effect systems have explicit effect polymorphism, consider for example this signature for \mintinline{koka}|map| in Koka \cite{leijen2014koka}.
\begin{minted}[fontsize=\small]{koka}
fun map(xs: list<a>, f: a -> e b): e list<b>
\end{minted}

In the capture calculus, we instead use subcapturing to allow higher-order functions to perform some effects.
In particular, we have a universal capture set, denoted \mintinline{scala}|{*}|, to which all capture sets are subcapturing.
As a syntactic sugar, we denote \mintinline{scala}|A => B| impure functions from \mintinline{scala}|A| to \mintinline{scala}|B|, i.e. \mintinline{scala}|{*} (A -> B)|, and \mintinline{scala}|A -> B| for pure functions --- those that are not capturing any capability.
The signature of the \mintinline{scala}|map| method can be expressed with this syntax as \mintinline{scala}|def map[B](f: A => B): List[B]|, \\ which is exactly the standard signature already present in the Scala standard library.
Hence, the capture calculus can represent effect polymorphism with very low or even non-existent syntactic overhead, and can therefore be retrofitted into existing code without requiring an entire rewrite.

The universal capability \mintinline{scala}|{*}| can also be used to prevent variables from being leaked from continuations.
Indeed, a particularity of \mintinline{scala}|{*}| is that generic types cannot be instantiated with a type whose capture set is universal.
Conceptually, all variable of a capture set of a type argument must be present in the current environment at the instantiation site, which is never the case for the universal capture set, otherwise we could mint capabilities out of thin air, defeating the point.

To illustrate how the universal capture set can be used to prevent capabilities from leaking, recall our \mintinline{scala}|withFile| example.
Let us write its signature in the following way:

\begin{minted}[fontsize=\small]{scala}
def withFile[T](p: String,
                f: ({*} OutputStream) => T): T
\end{minted}

Now, our faulty \mintinline{scala}|makeLogger| function is prevented at compile-time.
It instantiates \mintinline{scala}|T := {stream} (String => Unit)|, but the capability \mintinline{scala}|stream| is not in scope at the call site of \mintinline{scala}|withFile|, so we need to widen it to the next bigger capture set.
We have no other choice than to use \mintinline{scala}|{*}| which is forbidden by the restriction discussed above.

        \subsection{Capture Tunneling}

Consider a simple \mintinline{scala}|Pair| type.

\begin{minted}{scala}
class Pair[+T, +U](x: T, y: U) {
    def fst: T = x
    def snd: U = y
}
\end{minted}

Imagine that we want to bundle two impure values, one that can throw errors, \mintinline{scala}|sqrt: {err} (Double -> Double)| and one that has access to the filesystem to print strings, \\ \mintinline{scala}|log: {fs} (String -> Unit)|.
What should be the capture set of \mintinline{scala}|val p = Pair(sqrt, log)| ?
One option is to float the capture sets outwards, and take their union, i.e. \mintinline{scala}|{err,fs}|.
However, such a type wouldn't be precise enough, \mintinline{scala}|p.fst| would be capturing the \mintinline{scala}|fs| capability despite the fact that it does not access the filesystem.
Consider a method \mintinline{scala}|mapFirst| that maps the first element of the pair.
\begin{minted}[fontsize=\small]{scala}
def mapFirst[A, B, C](p: Pair[A, B],
                      f: A => C): Pair[C, B] =
    Pair(f(p.fst), p.snd)
\end{minted}
If capture sets were propagated, one would have to annotate the \mintinline{scala}|p| parameter with the universal capture set, and consequently the return type would also have to capture \mintinline{scala}|{*}|, making the result type intolerably imprecise.

For this reason, the capture set of generic type arguments is not propagated beyond the instantiation site, and in our example, the capture set of \mintinline{scala}|val p = Pair(sqrt, log)| would thus be empty.
Then, at the use site of the field, we reveal its sealed capture sets.
For instance, we would like to type the closure \mintinline[breaklines]{scala}|() => p.fst(0.0)| as \mintinline[breaklines]{scala}|{err} () -> Double|.
This behavior is dubbed \textit{capture tunneling}, it allows us to describe capture sets tightly, and avoid operations on generic types to gradually lose in precision.

        \subsection{Boxes}

\begin{anf-ccsubbox}

Note that, if type parameters were constrained to be pure (uncaptured), then the problem described in the previous section would not arise, since there would be no capture sets.
Obviously, requiring all type arguments to be pure would be overly restrictive.
For this purpose, \ccsubbox introduces the concept of \textit{boxes}.
Box ($\Box$) is a modality that hides the captured variables of its argument, allowing us to treat a captured type as pure.
The purpose of boxing is to enforce type arguments to be pure without losing in expressivity compared to \fsub.
More precisely, one can instantiate a generic type with an impure type $\tycapt{C}{R}$, by putting it into a box, denoted $\tybox{\tycapt{C}{R}}$.
At any point, we can box a term, as long as its captured variables are currently in scope.
Then, we have to explicitly tell when we want to reveal (or \textit{unbox}) the captured variables of a boxed term, when we want to use it, again provided that the context in which we unbox contains underlying capture set.
It is the mechanism that enables capture tunneling, and thus more precise capturing, particularly in the presence of generics.

Going back to our \mintinline{scala}|Pair| example, \mintinline{scala}|p| would have to be declared as following:
\begin{minted}[escapeinside=||,mathescape=true]{scala}
    val p: Pair[|$\Box$| {err} (Double -> Double),
                |$\Box$| {fs} (String -> Unit)] =
        Pair(|$\Box$| sqrt, |$\Box$| log)
\end{minted}
Then, one can recover the underling type behind the box using the unboxing syntax $\tmunbox{C}{\tm{x}}$.
\begin{minted}[escapeinside=||,mathescape=true]{scala}
    val sqrt: {err} (Double -> Double) =
      {err} |$\multimapinv$| p.fst
\end{minted}
This is only legal if the \mintinline{scala}|err| capability is in scope at the unboxing site, e.g. if the unbox happens inside a \mintinline{scala}|try|-block.

While the addition of boxes is a significant departure from \fsub, it turns out that boxing and unboxing can be inferred in many cases in practice, as shown in \cite{xu2023formalizing}, keeping the syntactic overhead low.

\end{anf-ccsubbox}

    \section{Metatheory}

We now focus on the metatheory of the capture calculus and how we approached its mechanization.
In particular, we compare the mechanizations of \cfsub with the one of \ccsubbox in the hope of motivating the concept of boxes, but this time from a metatheoretical perspective. 
We explain how the restriction on polymorphic types bounds to be pure leads to a much simpler theory.

        \subsection{Pure Types versus Pretypes}

In \cfsub, the syntax of types is defined by two mutual syntactic categories: types and pretypes.
Types correspond to pretypes prefixed by a capture set, or a type variable that will later be substituted by a type.
On the other hand, pretypes can be function types, (bounded) polymorphic types or top types that can be found in \fsub.
The only difference with \fsub is the dependent nature of function types, since return types can refer to the name of their argument in capture sets.

However, in \ccsubbox, type variables are now considered to belong in the same syntactic categories as function types, polymorphic types and top types.
This syntactic category is referred to as "pure types".
The only other addition is the boxed type which masks variable dependencies represented in the capture set of a type.

\begin{figure}[H]%
    \centering%
    \begin{minipage}[t]{0.5\columnwidth}%
        \begin{anf-cfsub}%
            \begin{align*}
                &\begin{syncat}{\ty{T}, \ty{U}}
                    \syncon{\tycapt{C}{R} \sep \tyvar{X}}
                \end{syncat} \\
                &\begin{syncat}{\prety{Q}, \prety{R}}
                    \syncon{\tytop} \\
                    \syncon{\tyarr{\var{x} \sym{:} \ty{T}}{U}} \\
                    \syncon{\tyall{\var{X} \sym{\sub} \ty{T}}{U}} \\
                \end{syncat}%
            \end{align*}%
        \end{anf-cfsub}%
    \end{minipage}%
    \begin{minipage}[t]{0.5\columnwidth}
        \begin{anf-ccsubbox}
            \begin{align*}
                &\begin{syncat}{\ty{T}, \ty{U}}
                    \syncon{\tycapt{C}{R} \sep \prety{R}}
                \end{syncat} \\
                &\begin{syncat}{\prety{Q}, \prety{R}}
                    \syncon{\prety{X} \sep \tytop \sep \tybox{T}} \\
                    \syncon{\tyarr{\var{x} \sym{:} \ty{T}}{U}} \\
                    \syncon{\tyall{\var{X} \sym{\sub} \prety{R}}{U}} \\
                \end{syncat}%
            \end{align*}%
        \end{anf-ccsubbox}%
    \end{minipage}%
    \caption{Raw syntax for type in \cfsub (left) vs \ccsubbox (right)}%
    \label{fig:types-cfsub-vs-ccsubbox}%
\end{figure}

\begin{anf-ccsubbox}
In the mechanized soundness proof of \cfsub, the raw syntax of types and pretypes is described by mutually inductive types, and we have mutually inductive judgements to ensure closedness, well-formedness, and subtyping.

However, by considering type variables as pure types, we break the assumption that type substitution is stable by the syntactic category, i.e. that if we substitute in a pure type, we will obtain a pure type, and similarly if we substitute in a type we will obtain a type.
Indeed, $\tyvar{X}$ is pure but $\lnsubstt{X}{\tycapt{C}{R}}{\tyvar{X}} = \tycapt{C}{R}$ is not.
Moreover, we now have an injection from pure types to types.

Therefore, the mutual encoding of syntax that was used in the mechanization of the soundness proof of \cfsub does not fit as well in \ccsubbox.
This duplicates a lot of the lemmas and requires mutual induction, for which Coq can be very slow to check for termination.
For the reasons above, we made the decision to encode the raw syntax of pure types and types in a single inductive type and just have closedness be expressed in a mutually inductive fashion.
This means that we have two judgements $\type{T}$ and $\pure{R}$ -- see Figure~\ref{fig:ccsubbox-type-pure} -- as well as an injection $\pure{R} \Rightarrow \type{\prety{R}}$.
This change requires extra attention to not mix up captured types and pure types, but results in a drastically shorter proof, which is also faster to check.
For example, pure types need to be augmented by an empty capture set to be considered as captured types.
\end{anf-ccsubbox}

\begin{figure}
    \centering
    \begin{anf-ccsubbox}
    \begin{mathpar}
        \ir{capt}{
            \forall \bvar{i} \in \mathbb{N} \;\; \bvar{i} \notin \cap{C}
        }{
            \capt{C}
        } \quad
        \ir{capt-type}{
            \capt{C} \and \pure{R}
        }{
            \type{\tycapt{C}{R}}
        } \\
        \ir{pure-type}{
            \pure{R}
        }{
            \type{\prety{R}}
        } \quad
        \ir{tvar-pure}{}{
            \pure{\tvar{X}}
        } \\
        \ir{top-pure}{}{
            \pure{\tytop}
        } \quad
        \ir{box-pure}{
            \type{T}
        }{
            \pure{\tybox{T}}
        } \\
        \ir{arr-pure}{
            \type{S} \and
            \forall x \notin \fvar{L} \;\; \type{\lnopen{0}{x}{T}}
        }{
            \pure{\tyarr{S}{T}}
        } \\
        \ir{all-pure}{
            \pure{R} \and
            \forall \tvar{X} \notin \fvar{L} \;\; \type{\lnopent{0}{X}{T}}
        }{
            \pure{\tyall{R}{T}}
        }
    \end{mathpar}
    \end{anf-ccsubbox}
    \caption{\type{} and \pure{} judgements of \ccsubbox}
    \label{fig:ccsubbox-type-pure}
    \noindent
\end{figure}

        \subsection{Well-formedness}

\begin{anf-cfsub}
In \cfsub, the well formedness judgement needs to take care of variance which was somewhat complicated to deal with in the mechanized proof.
Due to the type/pretype distinction, it is split into two mutually inductive types, which are defined by four-place relations $\wftypin{\Gamma}{A_+}{A_-}{T}$ and $\wfpretypin{\Gamma}{A_+}{A_-}{R}$ for types and pretypes respectively, meaning "in context $\ctx{\Gamma}$ with positive occurrences of variables in $A_+$ and negative occurrences of variables in $A_-$, type $\ty{T}$ or pretype $\prety{R}$ is well formed".
See Figure~\ref{fig:cfsub-wftyp} in the appendix for the complete set of rules, presented in a locally-nameless style as in the mechanization.
Then, the usual well-formedness judgements for types and pretypes are defined as follows:
\begin{align*}
\wftyp{\Gamma}{T}    &\coloneqq \wftypin{\Gamma}{\dom{\Gamma}}{\dom{\Gamma}}{T}   \\
\wfpretyp{\Gamma}{R} &\coloneqq \wfpretypin{\Gamma}{\dom{\Gamma}}{\dom{\Gamma}}{R} \\
\end{align*}
\end{anf-cfsub}

\vspace{-2em}

With the addition of boxes, the well-formedness judgement can be expressed in \ccsubbox in a way that is closer to the well-formedness judgement of \fsub.
Indeed, in \cfsub, type variables can occur in capture sets.
This way one can instantiate a type abstraction with a captured type $\ty{T}$, and all occurrences of the bound type $\tvar{X}$ will be replaced with the captured variables of $\ty{T}$, i.e. all variables contained in the capture sets in $\ty{T}$ that are not behind boxes.
This feature is the reason why we need to parametrize the well-formedness judgement by atom sets.
However, \ccsubbox restricts instantiation to pure types, which means capture sets no longer need to account for type variables.
Instead, one can instantiate a type abstraction with a boxed type and later use unboxing to recover the underlying captured type.
Moreover, the single inductive type describing the raw syntax enables us to describe well-formedness in a single inductive type.
The complete well-formedness judgement for \ccsubbox is defined in Figure~\ref{fig:ccsubbox-wftyp}.
These design decision further cut down the length proof by a large amount, and simplified the overall mechanization process.

\begin{figure}
    \centering
    \begin{anf-ccsubbox}
    \begin{mathpar}
        \ir{capt-wf}{
            \cap{C} \subseteq \dom{\Gamma} \cup \{\star\} \quad
            \wftyp{\Gamma}{R} \quad
            \pure{R}
        }{
            \wftyp{\Gamma}{\tycapt{C}{R}}
        } \\
        \ir{top-wf}{}{
            \wftyp{\Gamma}{\tytop}
        } \\
        \ir{tvar-wf}{
            \bsub{X}{T} \in \ctx{\Gamma}
        }{
            \wftyp{\Gamma}{\tyvar{X}}
        } \quad
        \ir{box-wf}{
            \wftyp{\Gamma}{\tycapt{C}{R}}
        }{
            \wftyp{\Gamma}{\tybox{\tycapt{C}{R}}}
        } \\
        \ir{fun-wf}{
            \wftyp{\Gamma}{S} \quad
            \left( \forall \fvar{x} \notin \fvar{L} \;\; \wftyp{\Gamma}{\lnopen{0}{x}{T}} \right)
        }{
            \wftyp{\Gamma}{\tyarr{S}{T}}
        } \\
        \ir{tfun-wf}{
            \wftyp{\Gamma}{\prety{R}} \quad
            \pure{R} \quad
            \left( \forall \tvar{X} \notin \fvar{L} \;\; \wftyp{\Gamma}{\lnopent{0}{X}{T}} \right)
        }{
            \wftyp{\Gamma}{\tyall{R}{T}}
        }
    \end{mathpar}
    \end{anf-ccsubbox}
    \vspace{-2em}
    \caption{Locally-nameless presentation of the well-formedness rules for types in \ccsubbox}
    \label{fig:ccsubbox-wftyp}
    \noindent
\end{figure}

        \subsection{Subtyping}

Another consequence of encoding the raw syntax of types in a non-mutually inductive manner is that subtyping for types and pure types can be merged into a single non-mutual inductive judgement, just like well-formedness.
The former subtyping judgements for types and pretypes in \cfsub was defined as in Figure~\ref{fig:cfsub-subtyp}.

In the mechanized proof for \ccsubbox, the new single subtyping judgement is described in Figure~\ref{fig:ccsubbox-subtyp}.

However, we encountered an issue when trying to prove the transitivity of subtyping.
Transitivity of subtyping is one of the only "high-level" lemmas that we prove using mutual induction, since it depends on whether the middle type is pure or not.
Given the structure of types and pretypes, Coq cannot check that a naive mutually inductive proof terminates, because of the injection from pure types to types.
Using a custom combined induction principle lets us prove the transitivity of subtyping in a way that Coq can check for termination.

        \subsection{Reduction rules}

The small-step semantics of \fsub in the proof of Aydemir et al. \cite{aydemir2008engineering} are specified using a standard binary relation on terms.
However, due to the MNF structure of \ccsubbox, we can not substitute a term for another in general, we instead rely on a three-state abstract machine, similar to the CEK machine \cite{felleisen1986control}.
During our adaptation of \fsub to MNF-\fsub, we took care of defining the small-step semantics in a similar way to that of the \ccsubbox semantics illustrated in Figure~\ref{fig:ccsubbox-paper-reductions}.

\begin{figure}[h]
    \centering
    \begin{anf-ccsubbox}
    \begin{mathpar}
        \ir{app}{
            x = \tmabs{T}{e} \in S \and
            y = v \in S
        }{
            \red{}{
                \mkstate{S}{E}{\tmapp{x}{y}}
            }{
                \mkstate{S}{E}{\lnopen{0}{y}{e}}
            }
        } \\
        \ir{tapp}{
            x = \tmtabs{\prety{R}}{e} \in S \and
            \pure{R}
        }{
            \red{}{
                \mkstate{S}{E}{\tmtapp{x}{\prety{R}}}
            }{
                \mkstate{S}{E}{\lnopent{0}{\prety{R}}{e}}
            }
        } \\
        \ir{open}{
            x = \tmbox{y} \in S \and
        }{
            \red{}{
                \mkstate{S}{E}{\tmunbox{C}{x}}
            }{
                \mkstate{S}{E}{\fvar{y}}
            }
        } \\
        \ir{rename}{
            x = v \in S
        }{
            \red{}{
                \mkstate{S}{\evalcons{e}{E}}{x}
            }{
                \mkstate{S}{E}{\lnopen{0}{\fvar{x}}{e}}
            }
        } \\
        \ir{lift}{}{
            \red{}{
                \mkstate{S}{\evalcons{e}{E}}{v}
            }{
                \mkstate{\storecons{S}{x}{v}}{E}{\lnopen{0}{\fvar{x}}{e}}
            }
        } \\
        \ir{let}{}{
            \red{}{
                \mkstate{S}{E}{\tmlet{e_1}{e_2}}
            }{
                \mkstate{S}{\evalcons{e_2}{E}}{e_1}
            }
        }
    \end{mathpar}
    \end{anf-ccsubbox}
    \vspace{-2em}
    \caption{Reduction rules for the \ccsubbox abstract machine used in the mechanization}
    \label{fig:ccsubbox-coq-reductions}
\end{figure}

\begin{anf-ccsubbox}
Working with holed contexts as in Figure~\ref{fig:ccsubbox-paper-reductions} is however cumbersome in mechanized proofs.
We therefore represent our expressions by an abstract machine state triplet $\mkstate{S}{E}{e}$ where $S$ is a list of value bindings representing the store context, and $E$ is a list of continuations representing the evaluation context.
Our small-step semantics for \ccsubbox is given in Figure~\ref{fig:ccsubbox-coq-reductions}.
Note that the only additional rule compared to the holed-context-based reductions in Figure~\ref{fig:ccsubbox-paper-reductions} is the (\textsc{Let}) rule which simply builds up the evaluation context if the current focused expression is a let-binding.

\end{anf-ccsubbox}

    \section{Conclusion}

The mechanization of the soundness proof of \ccsubbox has shown that the \boxcalc is a cleaner and simpler formulation of capture checking.
However, the restrictions needed to simplify the metatheory, namely forcing the user to box types before instantiating a polymorphic type, mean that we need a mechanism to infer those boxes to make the calculus accessible and seamlessly integrate with existing code, as described in \cite{xu2023formalizing}.
Nevertheless, the mechanization of \ccsubbox is almost half as long as the one of \cfsub, going from \texttilde 12k LOC to \texttilde 7k LOC.
Our proof is available on GitHub at \href{https://github.com/felko/ccsubbox}{https://github.com/felko/ccsubbox}.

    \section{Future Work}
        \subsection{Speeding Up the Proof Checking}

Judging the quality of a mechanized proof does not reduce to whether the proof assistant accepts it.
Specifically, we want our proofs to be fast to check and robust to changes in definitions.
However, these aspects are somehow in tension: robust proofs tend to use more automation which leads them to be shorter but also longer to verify.
Instead of writing all proof steps, we rely on tactics to do the tedious, overly formal work for us.
One such tactic is the \verb|pick fresh| tactic \cite{aydemir2008engineering} which gathers all variables in context and generates a fresh variable together with a proof that it is actually fresh (i.e. that it is not in the gathered set of atoms).
Then, we prove freshness goals of the form $x \notin ...$ using the \verb|fsetdec| tactic from the Coq standard library.
The \verb|fsetdec| tactic can sometimes take a few seconds to complete when there are lots of atoms in scope.
A possible improvement over the current status is to differentiate atoms by what they stand for, i.e. atoms representing term variables should not be confused with atoms standing for type variables.
One could parametrize the \verb|atom| type by a syntactic category (such as \verb|exp| or \verb|typ|) and specifying in the \verb|pick fresh| tactic when we want a fresh term variable or a fresh type variable.
By doing so, the gathered sets of atoms should be smaller and we can hope that the calls to \verb|fsetdec| will be faster.

        \subsection{Using Automation Libraries}

The current winner of the \poplmark challenge, in which programming language researchers compete to mechanize the soundness of \fsub in the least amount of code, is \href{https://www.ps.uni-saarland.de/autosubst/}{\autosubst} \cite{schafer2015autosubst}.
\autosubst is a Coq library tailored for automating the proof of substitution lemmas, which account for a large portion of the overall proof.

\begin{figure}[H]
    \centering
    \includegraphics[width=\columnwidth,keepaspectratio]{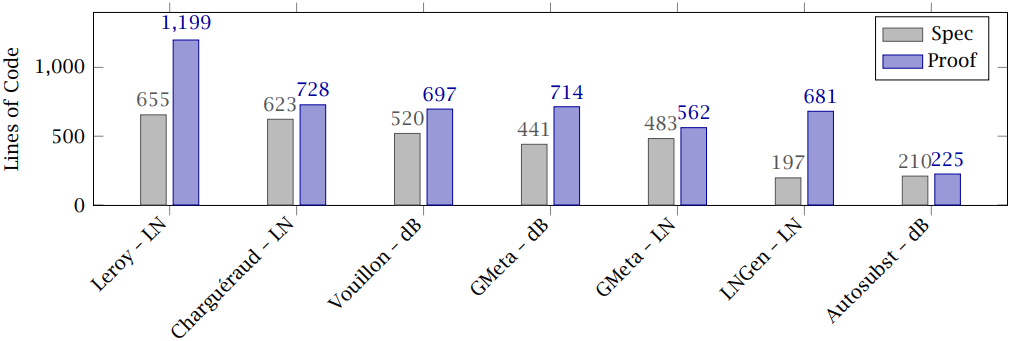}
    \caption{Comparison of \poplmark challenge submissions (image taken from \cite{schafer2015autosubst})}
    \label{fig:poplmark-comparison}
\end{figure}

\vspace{-1em}

Our current mechanization is based on Arthur Charguéraud's locally nameless proof \cite{chargueraud2012locally}, and only uses small tactic libraries, namely \textsc{LibTactics} and \textsc{TaktikZ}.
One could wonder if \autosubst could be applied to \ccsubbox and how small would the soundness proof be if we were to use it.
One major difference is that \autosubst uses De Bruijn indices instead of locally nameless.

\printbibliography

\newpage

\appendix

\section{Appendix}
    
\begin{figure}[h]
    \centering
    \begin{anf-cfsub}
    \begin{mathpar}
        \ir{capt-wf}{
            \cap{C} \subseteq \fvar{A_+} \and
            \wfpretypin{\Gamma}{A_+}{A_-}{U} \\\\
            \forall x_i \in \cap{C} \; \btyp{x_i}{S_i} \in \ctx{\Gamma}
        }{
            \wftypin{\Gamma}{A_+}{A_-}{\tycapt{C}{U}}
        } \\
        \ir{universe-wf}{
            \wfpretypin{\Gamma}{A_+}{A_-}{U}
        }{
            \wftypin{\Gamma}{A_+}{A_-}{\tycaptc{\star}{U}}
        } \\
        \ir{tvar-wf}{
            \bsub{X}{T} \in \ctx{\Gamma}
        }{
            \wfpretypin{\Gamma}{A_+}{A_-}{\tyvar{X}}
        } \\
        \ir{fun-wf}{
            \wftypin{\Gamma}{A_-}{A_+}{S} \\
            \left( \forall \fvar{x} \notin \fvar{L} \;\; \wftypin{\Gamma}{A_+ \sym{\cup} \sym{\{}\fvar{x}\sym{\}}}{A_-}{\lnopen{0}{x}{T}} \right)
        }{
            \wfpretypin{\Gamma}{A_+}{A_-}{\tyarr{S}{T}}
        } \\
        \ir{tfun-wf}{
            \wftypin{\Gamma}{A_-}{A_+}{S} \\
            \left( \forall \tvar{X} \notin \fvar{L} \;\; \wftypin{\Gamma}{A_+ \sym{\cup} \sym{\{}\tvar{X}\sym{\}}}{A_-}{\lnopent{0}{X}{T}} \right)
        }{
            \wfpretypin{\Gamma}{A_+}{A_-}{\tyall{S}{T}}
        } \\
        \ir{top-wf}{}{
            \wfpretypin{\Gamma}{A_+}{A_-}{\tytop}
        }%
    \end{mathpar}%
    \end{anf-cfsub}%
    \vspace{-2em}
    \caption{Locally-nameless presentation of well-formedness rules for types ($\vdash_T$) and pretypes ($\vdash_R$) in \cfsub}
    \label{fig:cfsub-wftyp}
\end{figure}

\begin{figure*}
    \centering
    \begin{anf-cfsub}
    \begin{mathpar}
        \ir{sub-refl-tvar}{
            \wfenv{\Gamma} \and
            \wftyp{\Gamma}{\tyvar{X}}
        }{
            \subtyp{\Gamma}{\tyvar{X}}{\tyvar{X}}
        } \qquad
        \ir{sub-trans-tvar}{
            \bsub{X}{S} \in \ctx{\Gamma} \and
            \subtyp{\Gamma}{S}{T}
        }{
            \subtyp{\Gamma}{\tyvar{X}}{T}
        } \\
        \ir{sub-capt}{
            \subcapt{\Gamma}{C_1}{C_2} \and
            \subpretyp{\Gamma}{R_1}{R_2}
        }{
            \subtyp{\Gamma}{\tycapt{C_1}{R_1}}{\tycapt{C_2}{R_2}}
        } \qquad
        \ir{sub-top}{
            \wfenv{\Gamma} \and
            \wfpretyp{\Gamma}{T}
        }{
            \subpretyp{\Gamma}{T}{\tytop}
        } \\
        \ir{sub-fun}{
            \subtyp{\Gamma}{S_2}{S_1} \and
            \wftyp{\Gamma}{S_1} \and
            \wftyp{\Gamma}{S_2} \\\\
            \left( \forall x \notin \fvar{L} \;\; \wftypin{\ctxtyp{\Gamma}{x}{S_1}}{\dom{E} \cup \{x\}}{\dom{E}}{\lnopenct{0}{\{x\}}{T_1}} \right) \\\\
            \left( \forall x \notin \fvar{L} \;\; \wftypin{\ctxtyp{\Gamma}{x}{S_2}}{\dom{E} \cup \{x\}}{\dom{E}}{\lnopenct{0}{\{x\}}{T_2}} \right) \\\\
            \left( \forall x \notin \fvar{L} \;\; \subtyp{\ctxtyp{\Gamma}{x}{S_2}}{\lnopenct{0}{\{x\}}{T_1}}{\lnopenct{0}{\{x\}}{T_2}} \right)
        }{
            \subpretyp{\Gamma}{\tyarr{S_1}{T_1}}{\tyarr{S_2}{T_2}}
        } \\
        \ir{sub-tfun}{
            \subtyp{\Gamma}{S_2}{S_1} \and
            \wftyp{\Gamma}{S_1} \and
            \wftyp{\Gamma}{S_2} \\\\
            \left( \forall X \notin \fvar{L} \;\; \wftypin{\ctxsub{\Gamma}{X}{S_1}}{\dom{E} \cup \{X\}}{\dom{E} \cup \{X\}}{\lnopenct{0}{\{X\}}{T_1}} \right) \\\\
            \left( \forall X \notin \fvar{L} \;\; \wftypin{\ctxsub{\Gamma}{X}{S_2}}{\dom{E} \cup \{X\}}{\dom{E} \cup \{X\}}{\lnopenct{0}{\{X\}}{T_2}} \right) \\\\
            \left( \forall X \notin \fvar{L} \;\; \subtyp{\ctxsub{\Gamma}{X}{S_2}}{\lnopenct{0}{\{X\}}{T_1}}{\lnopenct{0}{\{X\}}{T_2}} \right)
        }{
            \subpretyp{\Gamma}{\tyall{S_1}{T_1}}{\tyall{S_2}{T_2}}
        }
    \end{mathpar}
    \end{anf-cfsub}
    \caption{Locally-nameless presentation of the subtyping rules for types ($\vdash_T$) and pretypes ($\vdash_R$) in \cfsub}
    \label{fig:cfsub-subtyp}
\end{figure*}

\begin{figure*}
    \centering
    \begin{anf-ccsubbox}
    \begin{mathpar}
        \ir{sub-refl-tvar}{
            \wfenv{\Gamma} \and
            \wftyp{\Gamma}{\prety{X}}
        }{
            \subtyping{\Gamma}{\prety{X}}{\prety{X}}
        } \qquad
        \ir{sub-trans-tvar}{
            \bsub{X}{\prety{R}} \in \ctx{\Gamma} \and
            \subtyping{\Gamma}{\prety{R}}{\prety{S}}
        }{
            \subtyping{\Gamma}{\prety{X}}{\prety{S}}
        } \qquad
        \ir{sub-box}{
            \subtyping{\Gamma}{T_1}{T_2}
        }{
            \subtyping{\Gamma}{\tybox{T_1}}{\tybox{T_2}}
        } \\
        \ir{sub-capt}{
            \subcapt{\Gamma}{C_1}{C_2} \and
            \subtyping{\Gamma}{\prety{R_1}}{\prety{R_2}} \and
            \pure{R_1} \and
            \pure{R_2}
        }{
            \subtyping{\Gamma}{\tycapt{C_1}{R_1}}{\tycapt{C_2}{R_2}}
        } \qquad
        \ir{sub-top}{
            \wfenv{\Gamma} \and
            \wftyp{\Gamma}{\prety{R}} \and
            \pure{R}
        }{
            \subtyping{\Gamma}{\prety{R}}{\tytop}
        } \\
        \ir{sub-fun}{
            \subtyping{\Gamma}{\tycapt{C_2}{R_2}}{\tycapt{C_1}{S_1}} \and
            \left( \forall x \notin \fvar{L} \;\; \subtyping{\ctxtyp{\Gamma}{x}{\tycapt{C_2}{S_2}}}{\lnopenct{0}{\{x\}}{T_1}}{\lnopenct{0}{\{x\}}{T_2}} \right) \\
        }{
            \subtyping{\Gamma}{\tyarr{S_1}{T_1}}{\tyarr{S_2}{T_2}}
        } \\
        \ir{sub-tfun}{
            \subtyping{\Gamma}{\prety{R_2}}{\prety{R_1}} \and
            \left( \forall X \notin \fvar{L} \;\; \subtyping{\ctxsub{\Gamma}{X}{R_2}}{\lnopent{0}{\{X\}}{T_1}}{\lnopent{0}{\{X\}}{T_2}} \right) \\
        }{
            \subtyping{\Gamma}{\tyall{R_1}{T_1}}{\tyall{R_2}{T_2}}
        }
    \end{mathpar}
    \end{anf-ccsubbox}
    \caption{Locally-nameless presentation of the subtyping rules in \ccsubbox}
    \label{fig:ccsubbox-subtyp}
\end{figure*}

\begin{figure*}
    \centering
    \begin{standard-fsub}
        \begin{align*}
            \textrm{Store context:}
            \begin{syncat}{S}
                \syncon{[]}
                \syncon{\tmlet{x}{v}{S}}
            \end{syncat} \\
            \textrm{Evaluation context:}
            \begin{syncat}{E}
                \syncon{[]}
                \syncon{\tmlet{x}{E}{e}}
            \end{syncat} \\
        \end{align*}
        \vspace{-2em}
        \[
            \begin{array}{lllr}
                S[E[\tmapp{x}{y}]]    \; &\longrightarrow \; S[E[\lnsubst{z}{y}{e}]]  & \quad \textrm{if } S(x) = \tmabs{z}{T}{e}  & \quad \rulename{app}    \\
                S[E[\tmtapp{x}{T}]]   \; &\longrightarrow \; S[E[\lnsubstt{X}{T}{e}]] & \quad \textrm{if } S(x) = \tmtabs{X}{T}{e} & \quad \rulename{tapp}   \\
                S[E[C \multimapinv x] \; &\longrightarrow \; S[E[y]]    & \quad \textrm{if } S(x)  = \square y   & \quad \rulename{open} \\
                S[E[\tmlet{x}{y}{e}]] \; &\longrightarrow \; S[E[\lnsubst{x}{y}{e}]]  & \quad                                      & \quad \rulename{rename} \\
                S[E[\tmlet{x}{y}{v}]] \; &\longrightarrow \; S[\tmlet{x}{v}{E[e]}]    & \quad \textrm{if } E \neq []               & \quad \rulename{lift} \\
            \end{array}
        \]
    \end{standard-fsub}
    \caption{Reduction rules for \ccsubbox as defined in \cite{odersky2022scoped}}
    \label{fig:ccsubbox-paper-reductions}
\end{figure*}

\end{document}